\documentclass[aps,twocolumn,showpacs,amsmath,amssymb,superscriptaddress,floatfix]{revtex4}
\usepackage{hyperref}
\usepackage{graphicx}
\usepackage{hypernat}

\def\etal{{\em{et al.}}}

\begin{document}

\title{Electronic Structure and Spectra of CuO}

\author{C. E. Ekuma}
\altaffiliation{Electronic Address: cekuma1@lsu.edu}
\affiliation{Department of Physics \& Astronomy, Louisiana State University,
Baton Rouge, LA 70803, USA}
\affiliation{Center for Computation and Technology, Louisiana State University, Baton Rouge, LA 70803, USA}

\author{V. I. Anisimov}
\altaffiliation{Electronic Address: via@imp.uran.ru}
\affiliation{Institute of Metal Physics, Russian Academy of Sciences, 620990 Yekaterinburg, Russia}
\affiliation{Ural Federal University, 620990 Yekaterinburg, Russia}

\author{J. Moreno}
\altaffiliation{Electronic Address: moreno@lsu.edu}
\affiliation{Department of Physics \& Astronomy, Louisiana State University,
Baton Rouge, LA 70803, USA}
\affiliation{Center for Computation and Technology, Louisiana State University, Baton Rouge, LA 70803, USA}

\author{M. Jarrell}
\altaffiliation{Electronic Address: jarrellphysics@gmail.com}
\affiliation{Department of Physics \& Astronomy, Louisiana State University,
Baton Rouge, LA 70803, USA}
\affiliation{Center for Computation and Technology, Louisiana State University, Baton Rouge, LA 70803, USA}


\begin{abstract}
\noindent We report the
electronic structure of monoclinic CuO as obtained
from first principles calculations utilizing density functional theory plus 
effective Coulomb interaction (DFT + U) method. 
In contrast to standard DFT calculations 
taking into account electronic correlations in DFT + U gave antiferromagnetic insulator with 
energy gap and magnetic moment values in good agreement with experimental data. 
The electronic states around the Fermi level are formed by partially filled 
Cu 3$d_{x^2-y^2}$ orbitals with significant admixture of O 2$p$ states. 
Theoretical spectra are calculated using DFT + U electronic structure method and their 
comparison with experimental photoemission and optical spectra show very good agreement. 
\end{abstract}

\maketitle

\section{Introduction}
\label{sec:intro}
The cupric oxide (CuO) system has been studied for decades both for fundamental understanding and applied 
reasons. It is generally recognized as the prototype material of a broad family of 
the strongly correlated (SC) oxides. \cite{Ghijsen1988} Although the local environment of Cu in the Cu-O planes 
are strongly distorted, the planes share many commonalities to those thought to be responsible 
for superconductivity in the layered cuprate systems.\cite{0295-5075-7-6-011,PhysRevLett.85.5170} 
A consequence of this, a good understanding of 
the electronic structure of CuO will play a key role in understanding and developing models 
for the normal state behavior of the high temperature superconducting cuprates. 
High-temperature superconductivity was 
discovered in the copper oxide perovskites as early as 1986, \cite{RevModPhys.60.585,Bednorz1986} 
but its origin and mechanism 
are still under intensive debate. 

CuO is an exceptional member of the generally, rocksalt family as it deviates 
both structurally and electronically from others as one traverses the periodic table from MnO to CuO. 
Unlike other members of the 3$d$ transition oxides (TMO) which crystallizes in the cubic 
rocksalt structure (with possible rhombohedral distortions),\cite{PhysRevB.5.290,PhysRevB.30.4734,PhysRevB.76.054417} 
Tenorite (CuO) crystallizes in the lower symmetry monoclinic 
(C$^6_{2h}$) crystal strucutre. Also, Unlike other antiferromagnetic (AFM) semiconductors, 
with known disordered paramagnetic character above 
the N\'{e}el temperature, CuO behaves as a 1D antiferromagnet \cite{PhysRevB.68.224433} 
with strong antiferromagnetic ordering 
especially along the (10$\tilde{1}$) direction prevailing even above the N\'{e}el temperature of 231 K. 
A 3D collinear antiferromagnetic order has been reported below 213 K, \cite{PhysRevB.39.4343} while between 
213 K $<$ T$_N$ $\lessapprox$ 231 K, it is reported to have a 3D non-collinear antiferromagnetic order. 
\cite{0022-3719-21-15-023} 
Also, CuO has substantially lower than expected 
N\'{e}el temperatue T$_N$ $\approx$ 230 K, following a simple linear extrapolation of the trend of other TMOs across 
the periodic table. However, like other TMOs, CuO is an antiferromagnetic insulator. \cite{PhysRevB.5.290,
Anisimov1997,PhysRevB.44.943,PhysRevB.76.054417}

Aside the fundamental importance of CuO in understanding the
properties of high-temperature superconductivity, it has other important 
practical technological applications. CuO has found practical applications 
in areas such as gas sensor,~\cite{Poizot2000} solar cells and 
photovoltaics,~\cite{Ray2001,Ben2009} catalysts,~\cite{A801595C} 
varistors,~\cite{Jiang1998} electrode in lithium ion batteries,~\cite{Ben2009,Morales2009} and 
magnetic storage media. \cite{doi:10.1021/cm000166z} 
Also, studies on CuO have shown strong dependence of its properties on quantum 
size effects,\cite{PhysRevB.61.11093,Rehman2011} and has recently been 
shown to exhibit multiferroicity at T$_c$ $\sim$ 
230 K.~\cite{Kimura2008,PhysRevLett.108.187205} 

Experimentally, CuO has a monoclinic crystal structure with C2/c symmetry.
\cite{Samokhvalov1989,ICSD2011} It has 
eight (8) formula units per magnetic unit cell. It is further reported to be 
a $p$-type semiconductor with band gap of 1.0 -- 1.9 eV~\cite{0953-8984-24-17-175002,
PhysRevB.52.1433,Ray2001,Madelung2006} and local moment per Cu atom of 
$\sim$ 0.7 $\mu_B$. \cite{PhysRevB.39.4343,0953-8984-4-23-009} 
Standard density functional theory (DFT)
with local exchange-correlation functionals (see for e.g., Refs.\onlinecite{PhysRevB.40.7684,
Kanagaraj,PhysRevB.87.115111}) generally predict a nonmagnetic ground state with metallic 
character instead of the well-known semiconducting ground state.
The failure of standard DFT 
to obtain the correct electronic properties of CuO should be understood from its intrinsic 
nature (inability to treat electron-electron interactions in the so-called 
correlated systems). Improvements in first-principle theories based on density functional theory (DFT) 
plus screened Coulomb interaction (U) (DFT + U) as proposed, developed, and utilized by 
Anisimov and co-workers~\cite{PhysRevB.44.943,PhysRevB.49.14211,
PhysRevB.52.R5467,Anisimov1997} has remedied this situation. While this method has been used to 
study the band structure of CuO (see for e.g., Refs.~\cite{PhysRevB.73.235206,Anisimov1997,PhysRevB.44.943}), 
we are not aware of any optical study of CuO utilizing this method.  

There have been many theoretical\cite{0953-8984-22-4-045502,Anisimov1997,PhysRevLett.108.187205,
PhysRevB.73.235206,PhysRevB.79.195122,PhysRevLett.106.026401,PhysRevLett.106.257601,PhysRevB.84.115108} 
and experimental studies\cite{Kimura2008,PhysRevB.39.4343,Gizhevski2006,0953-8984-17-3-009,
0953-8984-24-17-175002,Samokhvalov1989,0953-8984-11-26-306} of CuO. 
The purpose of this paper is to present optical properties of CuO 
based on the modern electronic structure that 
yield electronic properties of the strongly correlated systems in agreement with experiments,
\cite{PhysRevB.73.235206,Anisimov1997,PhysRevB.44.943}
in contrast to standard density functionals.
Therefore, enables direct, quantitative comparisons of band structures and
optical properties with experiment, without any adjustments, such as scaling the magnitude of the absorption
or applying scissors operators to fix the gap.
We extensively discuss the results in relation to
experimental data. We hope that they will motivate future
experimental investigations of the band structure of CuO
particularly using optical measurements and photoemission.

The rest of this article is organized as follows. After this introduction in 
section ~\ref{sec:intro}, the computational methods and details are given in 
section ~\ref{sec:formalism}. The results of our self-consistent calculations are presented and 
discussed in section~\ref{sec:results}. We will then conclude in section~\ref{sec:conclusion}. 

\section{Method and Computational Details}
\label{sec:formalism}
One difficulty in the computation of material properties is that the band gaps and related
properties of most materials are generally underestimated by the standard density functional
theory (DFT) approximations. To avoid this, we utilized the density functional theory (DFT) plus 
the effective Coulomb interaction (U)  (DFT + U) formalism using the general potential linearized augmented
planewave (LAPW) method, \cite{singh2006} as implemented in the WIEN2k code. \cite{Blaha2001} 
Unlike other DFT + U computations for the electronic properties of CuO \cite{PhysRevLett.108.187205,
PhysRevB.73.235206,PhysRevB.84.115108} that utilized 
the effective U-value (U$_{eff}$) calculated for e.g., CaCuO$_2$\cite{PhysRevB.44.943} and 
La$_2$CuO$_4$,~\cite{PhysRevB.71.035105} in our case, 
we have self-consistently computed the U$_{eff}$ unique to CuO using the 
constrained DFT + U scheme
of Anisimov \etal~\cite{Anisimov1991} as implemented by Madsen \etal~\cite{Madsen2005} in WIEN2k code. \cite{Blaha2001} 
With this approach,
the effective Coulomb interaction (U$_{eff}$) on the Cu $d$ state is calculated self-consistently.
The computed value of U$_{eff}$ is 
7.14 eV. The DFT part of the computation
utilized the Perdew-Burke-Ernzerhof generalized gradient approximation (PBE-GGA).\cite{Perdew1996}

Aside from the use of the DFT + U formalism,
the other details of our calculations
are standard. We used well converged basis sets with dense Brillouin zone
samplings which is needed for the optical properties.
For this purpose we used a uniform 48$\times$48$\times$48 grid.
The LAPW sphere radii are 1.92 and 1.71 bohr for Cu and O. 
We utilized the room temperature experimental monoclinic crystal strucuture with lattice
parameter, a = 4.6837 \AA{}, b = 3.4226 \AA{}, and c = 5.1288 \AA{}. \cite{ICSD2011,Madelung2006} 
All calculations are performed relativistically. We carried out several 
sets of self-consistent calculations using different magnetic configurations. 
All attempts to obtain a ferromagnetic or non-magnetic solutions always led to 
an antiferromagnetic ground state.\cite{PhysRevB.5.290,
Anisimov1997,PhysRevB.44.943,PhysRevB.76.054417} This is in agreement with other results that 
show that CuO is an antiferromagnetic insulator. Since spin-orbit coupling (SOC) is important for CuO, 
reported results are for DFT + U + SOC. 
As mentioned, we apply no scissors operators or other adjustments.

\section{Results and Discussion}
\label{sec:results}

\subsection{Band Structure}
\begin{figure}
\includegraphics[trim = 0mm 50mm 5mm 10mm,width=1\columnwidth,clip=true]{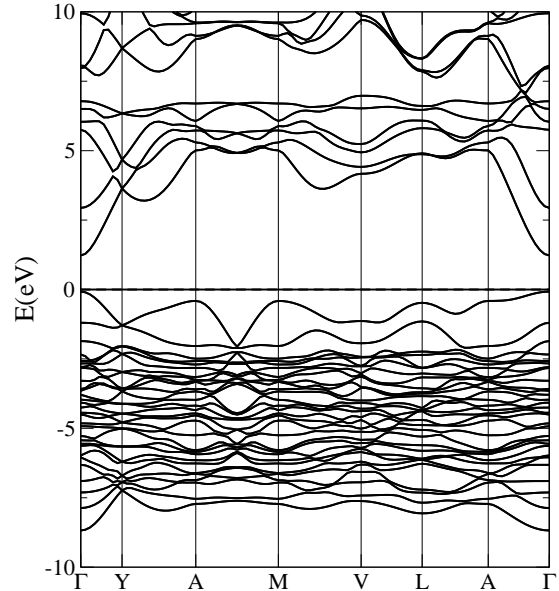}
\caption{(Color online) The calculated band structure of AFM CuO. The horizontal straight line is the position of 
the Fermi energy (E$_\mathrm{F}$) which has been set equal to zero.} 
\label{fig:CuO_bands}
\end{figure}

We begin with the calculated band structure in relation to
experimental data.
Our calculated band structure
for CuO is given in Fig.~\ref{fig:CuO_bands}. 
The band structure is qualitatively similar to those reported
previously, but there are quantitative differences resulting from
the use of the DFT + U functional, and these are important.

First of all, we note the values of the band gap.
We find band gap $E_g$= 1.25 eV (1.251 eV).
The calculated band gap is in accord with experimentally reported 
ambient temperature band gap value in the range 1.0 to 1.9 eV.~\cite{0953-8984-24-17-175002,
PhysRevB.52.1433,Ray2001,Madelung2006} The hybrid functional results of Rocquefelte \etal \cite{0953-8984-22-4-045502} 
reported a band gap of 1.42 eV (for mixing ratio $\alpha$ = 0.15) and 
2.4 eV (for mixing ratio $\alpha >$  0.15). 
The Green function plus screened Coulomb interation (GW) and GW plus on-site potential 
(GW + V$_d$) work of Lany\cite{PhysRevB.87.085112} reported values of 2.42 and 1.40 eV, respectively.

We did calculations as a function of the screen Coulomb values around the self-consistently 
determined value of 7.14 eV ($U_{eff}$ = 5,6,7,8 eV). Utilizing these $U_{eff}$ values to 
carry-out self-consistent computations, we only noticed a change of 0.12 eV in the 
band gap, we conclude that this is too little to cause any significant change in the band gap of CuO. 
We also carefully checked the dependence of the gap on various computational
parameters, such as energy cut-offs, energy window size for the SOC
calculation, LAPW sphere radii and Brillouin zone sampling.
We find that at worst, the errors related to these are less than 0.08 eV in the gap. In both 
cases, the position of the valence band maximum and the conduction band minimum didn't change.

The calculated magnetic moment per Cu atom (M$_{Cu}$)
(in the units of $\mu_B$) is 0.68 $\mu_B$ in good agreement with the experimental reported value of 
$\sim$ 0.7 $\mu_B$. \cite{PhysRevB.39.4343,0953-8984-4-23-009} We note that the 
magnetic moment per oxygen atom (M$_O$) is significant. It is 0.18 $\mu_B$
in good agreement with experimental value of 0.14 $\mu_B$.\cite{0022-3719-21-15-023} We note that 
the hybrid functional results of Rocquefelte \etal \cite{0953-8984-22-4-045502} reported M$_{Cu}$ 
and M$_O$ value of 0.65 and 0.12 $\mu_B$ (for mixing parameter $\alpha$ = 0.15), while 
M$_{Cu}$ = 0.74 and M$_O$ = 0.09 $\mu_B$ (for $\alpha >$  0.15), respectively.
 
\subsection{Comparison with Photoemission Spectra}
Photoemission spectra (PES) experiments provide a direct measure of 
the electronic structure of the occupied states. 
There have been PES experiments for CuO,
with which we can compare with.
\cite{Ghijsen1988,Shen1990,Thuler1982,PhysRevLett.78.1126,0953-8984-11-26-306}

\begin{figure}
\includegraphics[trim = 0mm 0mm 0mm 0mm,width=1\columnwidth,clip=true]{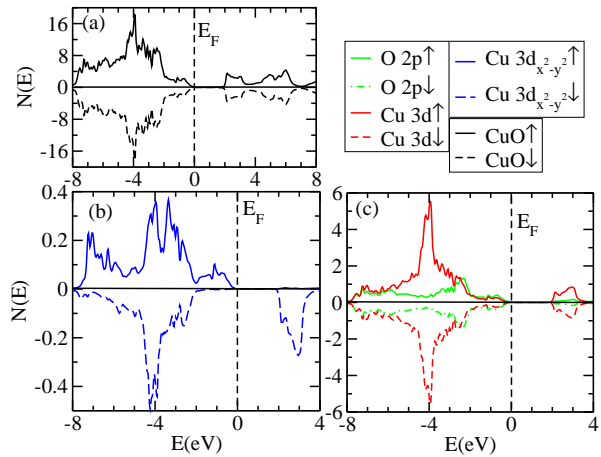}
\caption{(Color online) (a) The calculated total density of states of AFM CuO. (b) The calculated projected density of 
states of the $x^2-y^2$ Cu 3$d$ states. (c) The 
calculated projected density of states of the O 2$p$ and Cu 3$d$ states. Figure~\ref{fig:CuO_dos}(b) has been 
plotted using the local coordinate system 
centered on Cu where x and y axes are directed approximately to four neighboring oxygen.
The vertical straight line is the position of 
the Fermi energy (E$_\mathrm{F}$) which has been set equal to zero.} 
\label{fig:CuO_dos}
\end{figure}

\begin{figure}
\includegraphics[width=1\columnwidth,clip=true]{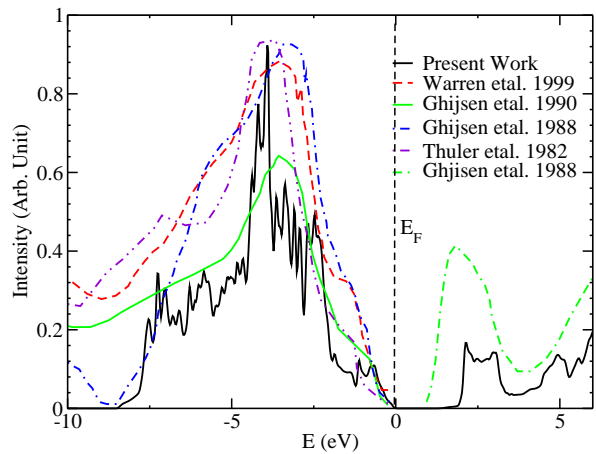}
\caption{(Color online) Calculated AFM CuO valence and low energy conduction band spectra without broadening as 
compared with experimental spectra.\cite{0953-8984-11-26-306,Ghijsen1988,Ghijsen1990,Thuler1982}. 
The experimental data are: the PES study of Warren \etal \cite{0953-8984-11-26-306} 
(plot key Warren etal. 1999), the resonant PES study 
of Ghijsen \etal~\cite{Ghijsen1990} (plot key Ghijsen etal. 1990), the 
PES study of Ghijsen \etal~\cite{Ghijsen1988}(measured with 
He II radiation) (plot key Ghijsen etal. 1998), and the PES study 
of Thuler \etal~\cite{Thuler1982} (plot key Thuler etal. 1982). 
The vertical straight line is the position of the Fermi energy which has been set equal to zero.} 
\label{fig:cuoPES}
\end{figure}

In Fig.~\ref{fig:CuO_dos}, we show the computed density of states (DOS) (Fig~\ref{fig:CuO_dos}(a)) and the 
partial density of states (Figs.~\ref{fig:CuO_dos}(b) to \ref{fig:CuO_dos}(c)) of CuO for both the spin majority and 
minority channels. 
While the bands are derived from a strong hybridization between Cu 3$d$ and O 2$p$ states, the conduction band 
(up to $\sim$3.56 eV) in the proximity of the Fermi level are formed mainly by the Cu 3$d$ 
(Cu 3$d_{x^2-y^2}\downarrow$) states (c.f. Fig~\ref{fig:CuO_dos}(b)). This corresponds to a case where 
there is a hole in the Cu 3$d_{x^2-y^2}$ states, the so-called 3d$^9$ configuration (we have used  the 
local coordinate system 
centered on Cu where x and y axes are directed approximately to four neighboring oxygen). In one of the 
earliest ab-initio study of orbital decomposition in CuO,\cite{PhysRevB.44.943} it was noted that the minority 
spin (Cu 3$d_{x^2-y^2}\downarrow$) states contribute strongly to the conduction band minimum and the 
(majority spins) Cu 3$d_{x^2-y^2} \uparrow$ and the O 2$p$ states 
constitute the bands in the valence band maximum. While 
the bands in the proximity of the Fermi energy (E$_F$) are predominantly Cu 3$d_{x^2-y^2}\downarrow$, 
there is a significant contribution from the O 2$p$ states. As it is evident from 
Figs.~\ref{fig:CuO_dos}(b) and (c), the spin polarization of the states around the Fermi energy arise entirely 
from Cu 3$d_{x^2-y^2}$ states.  
One can reconcile the various results that claim the non-detection 
of the 3$d^9$ configuration (see for e.g., Ref.~\onlinecite{0965-0393-14-2-006}) with those that 
actually see it, by noting that the 3$d^9$ configuration is very complicated
\cite{0965-0393-14-2-006,0022-3719-21-15-023,PhysRevB.56.12818} and observing it both 
in experiments and computations require very careful handling of the orbital decomposition. 
The most prominent feature in the conduction band are broad peaks at 2.13, 3.02, 4.97, and 5.98 eV. There 
are shallow minima around 3.62 $\pm$ 0.33 and 5.29 eV.

The states around the valence band maximum are predominantly of the O 2$p$ states hybridizing with 
the Cd $d$ (only the Cu 3$d_{x^2-y^2} \uparrow$) 
states in agreement with PES result of Warren \etal\cite{0953-8984-11-26-306}, resonant and X-ray 
photoemission spectroscopy (XPS) results of Shen \etal \cite{Shen1990}, and the X-ray photoemission 
spectroscopy results of Ghijsen \etal \cite{Ghijsen1988}. 
One can thus conclude that the states in the proximity of the Fermi energy 
are basically huge complex band of entirely the Cu 3$d_{x^2-y^2}$ and O 2$p$ states. The most prominent structure 
in the valence states is a peak at $\sim$ 3.93 eV. This peak is exclusively derived from the Cu 3$d$ states.

Figure~\ref{fig:cuoPES} shows the comparison of our computed DOS with experiments. 
We reiterate that our spectra were not shifted to coincide with experiment contrary to what
is often done in such comparisons with experiments (e.g. Ref. \onlinecite{PhysRevB.40.7684}). 
As can be seen from Fig.~\ref{fig:cuoPES}, the overall valence band spectra 
agree well with experimental ones. In particular, the agreement between our computed results and 
experimental spectra is reasonably good, both in terms of the relative intensities of the resonances 
and their binding energy positions. We note that there may be more features in our computed spectra 
as no broadening is used. Small deviation from experiment can be attributed to the fact that 
the intensity of photoemission spectra (PES) depend very sensitively on the photoionization across sections 
of the atomic sublevels. This deviation can even be seen between different experimental results. 
 
The features from 0 to $\approx$ 8.0 eV below E$_F$ are predominantly due to Cu 3$d$ and O 2$p$ states in basic agreement 
with the x-ray photoelectron spectroscopy (XPS) and reflection electron energy-loss spectroscopy (REELS) results of 
Tahir \etal~\cite{0953-8984-24-17-175002}. We calculate the valence band width to be $\approx$8.41 eV 
and the position of the maximum of the valence band
is locate at $\approx$3.83 eV in basic agreement with experimental ones with reported band width of 
7.8 -- 8.5 eV and the position of the valence band maximum located at 3 -- 4 eV.~\cite{0953-8984-11-26-306,
Ghijsen1988,Ghijsen1990,Thuler1982,Shen1990} Taking a closer look at the valence bands in the proximity of the 
Fermi energy, we find that these states are highly localized. We 
predict peaks at $\sim$ 0.66 eV and 1.10 eV. This feature is reminiscent of the 
so-called $^{1}A_{1g}$ singlet.\cite{Shen1990,Ghijsen1990,PhysRevB.41.288,
0953-8984-11-26-306} This singlet is formed due to the 
hybridization of Cu 3$d$ (mainly the Cu 3d$_{x^2-y^2}$) and O 2$p$ states.\cite{PhysRevB.38.6650} 
It is the so-called d$^9$\underline{L}$^1$ (where $\underline{L}$ is the ligand oxygen hole) 
final state with one hole in the 3d$_{x^2-y^2}$ orbital and the other in 
an O 2$p$ orbital.\cite{PhysRevB.41.288,Ghijsen1988}    
We predict antiresonance dip around 1.23 $\pm$ 0.12 eV in basic agreement with experiment~\cite{0953-8984-11-26-306} 
and a peak at 2.44 $\pm$ 0.21 eV. There are cluster of shoulders 
at 3.09 $\pm$0.22 and 5.2 $\pm$0.32 eV. The ultraviolet photoemission data of Ghijsen\etal~\cite{Ghijsen1988,Ghijsen1990} 
reported features at 1.23 $\pm$0.11, 3.10 $\pm$0.11, and 5.50 $\pm$0.40 eV.  

The low energy conduction band spectra show good agreement with the inverse photoemission 
(Bremsstrahlung isochromat spectroscopy) spectra experiment of Ghijsen \etal~\cite{Ghijsen1988}. Our computation 
predicted peaks at 2.11, 2.98 $\pm$ 0.20, and 4.97 eV. 

\subsection{Optical Properties and Comparison with Experiment}
Optical spectroscopy,
while less direct than PES, provides detailed
information about the electronic structure and has the advantage of
being a true bulk probe.

\begin{figure}
\includegraphics[trim = 0mm 20mm 5mm 0mm,width=1\columnwidth,clip=true]{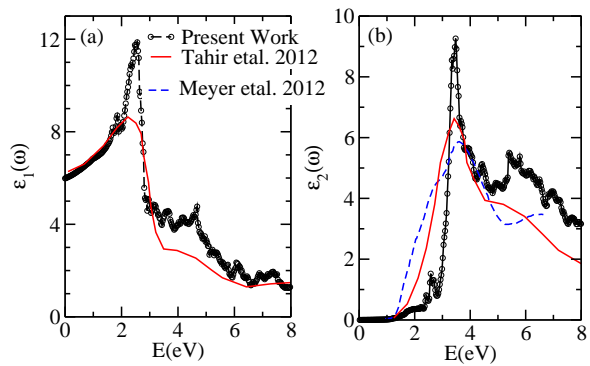}
\caption{(Color online) (a) Calculated, dispersive part, $\varepsilon_{1} (\omega)$, of the dielectric function 
of AFM CuO. 
(b) Calculated, absorptive part, $\varepsilon_{2} (\omega)$, of the dielectric function 
of CuO. Both $\varepsilon_{1} (\omega)$ and $\varepsilon_{2} (\omega)$ are compared to the experimental spectra of 
Tahir \etal~\cite{0953-8984-24-17-175002}. Note, both spectra have not been broadened as such, may have more structure 
than the experimental ones.} 
\label{fig:epsilon_cuo}
\end{figure}

The optical properties of CuO in both film and bulk
have been studied using several experimental methods.
\cite{0953-8984-24-17-175002,Harold1966,PhysRevB.52.1433,Jundale,Chen2009927,Balamurugan200190} 
Most of the
studies in the visible and ultraviolet have been near-normal incidence 
reflectivity measurements.\cite{Harold1966,PhysRevB.52.1433,Jundale,Chen2009927}
These methods have certain potential sources of
errors. In general, they involve multiple measurements
to obtain the dielectric constants and/or 
Kramers-Kronig (KK) analysis, which can
suffer from uncertainty in the absolute amplitude.
This is associated with extrapolations necessary for the
KK transformation of experimental reflectivity spectra. 

Spectroscopic ellipsometry (SE)
is one of the parallel measurement techniques that avoids
these problems 
in measuring the optical constants of solids.\cite{Aspnes1985}
The big advantage of the SE technique, and other 
parallel measurement techniques,
is that both the real and imaginary part of the complex dielectric function
of a material may be obtained directly as a function
of wavelength without requiring multiple measurements or
KK analysis. One of the earliest SE measurements for
CuO are those of
Ito and co-workers~\cite{Ito1998} in the 
1.2--5.0 eV photon-energy range at room temperature, Meyer and 
co-workers,~\cite{PSSB:PSSB201248128} and Nomerovannaya and co-workers.~\cite{Nomerovannaya}

As previously noted, we calculated
optical properties based on our DFT + U electronic structure,
with no adjustment.
This was done
using the optical properties package of the WIEN2k code.
While it is conventional to plot calculated optical data
with a broadening added to mimic experimental data, we instead
show results with no added broadening in order to show more clearly
the features in the calculated spectra.
Our calculated dispersive part of the dielectric function,
$\varepsilon_{1} (\omega)$ and the absorptive part, $\varepsilon_{2} (\omega)$ are shown in 
Fig.~\ref{fig:epsilon_cuo} in comparison with the experimental 
results of Tahir and Tougaard~\cite{0953-8984-24-17-175002} and Meyer \etal~\cite{PSSB:PSSB201248128}. 
As can be seen, our computed spectra for 
$\varepsilon_{1} (\omega)$ (c.f. Fig.~\ref{fig:epsilon_cuo}(a)) and 
$\varepsilon_{2} (\omega)$ (c.f. Fig.~\ref{fig:epsilon_cuo}(b)), are in good 
agreement with experiment. 

There is no yet rigorous ab-initio optical property study of CuO especially, 
taking into account the strong electron-electron correlations; as such, this 
study serves as a baseline for comparing with future experiment and theory.

Figure~\ref{fig:epsilon_cuo}(a) shows our 
calculated dispersive part of the dielectric function,
$\varepsilon_{1} (\omega)$ in comparison with the data of Tahir \etal~\cite{0953-8984-24-17-175002} 
and Meyer \etal~\cite{PSSB:PSSB201248128}.
The main experimental features
in CuO \cite{0953-8984-24-17-175002} 
are reproduced in our results.
The main feature is a shoulder
at 1.82 $\pm$ 0.2 eV, followed by a steep 
rise. We predict cluster of small shoulders at 3.05 $\pm$ 0.42 eV. Our data show 
a dip at $\sim$ 3.82 eV. Experiment show 
a similar dip at $\sim$ 3.68 eV.\cite{0953-8984-24-17-175002} 
The SE data of Ito \etal~\cite{Ito1998} show features at $\sim$ 1.6, 2.0, 2.6, and 3.4 eV.~\cite{Ito1998} 
We predict a shoulder at $\sim$ 6.62 $\pm$ 0.21 eV followed by a steep 
decrease slowly towards almost zero at higher energies. 
Our computed $\varepsilon(\infty)$ = $\Re (\varepsilon(0))$ is 6.12.
The infrared spectroscopy data of Kuz'menko \etal~\cite{PhysRevB.63.094303} reported 
$\varepsilon(\infty)$ of 5.9 -- 7.8 (average value is 6.60), 
the SE data of Ito \etal~\cite{Ito1998} reported a value of 6.45, and the 
polarized reflectance of a single crystal of CuO study of 
Homes \etal~\cite{PhysRevB.51.3140} reported a value of 6.2 -- 6.3. We note that 
the GW + V$_d$ results of Lany~\cite{PhysRevB.87.085112} reported static electronic 
dielectric constant of 7.9.  
The maximum amplitude of $\varepsilon_{1} (\omega)$ is 11.79 $\pm$0.15
at 2.52 $\pm$0.13 eV.    

Figure~\ref{fig:epsilon_cuo}(b) shows our 
calculated absorptive part of the dielectric function,
$\varepsilon_{2} (\omega)$ in comparison with the experiments.~\cite{0953-8984-24-17-175002}. 
We found a very small kink around 1.51 eV and then, shoulders around 2.40 and 2.61 eV,
followed by other sets of close dips around 2.30, 2.53, and 2.82 eV; after which, the 
spectra increased with a steep rise in energy. 
The maximum of $\varepsilon_{2} (\omega)$ is $\sim$8.96 at 
energy of 3.57 eV. The experimental data of Meyer \etal~\cite{PSSB:PSSB201248128} show features at 
1.66, 2.07, 2.68 eV with a maximum at 3.46 eV in good agreement with our computed results. 
We also find shoulders at 3.86, 4.45, and 5.41 eV; and cluster of dips at 
4.24, and 4.85 $\pm$ 0.32 eV.  

Thus there is a very close correspondence between our
first principles results,
which represent standard band structure descriptions,
and both optical data from SE and PES
measurements.

\section{Conclusion}
We have performed self-consistent DFT + U to study the electronic and optical 
properties of monoclinic CuO. Our computations show that 
there is significant Cu-$d$ and O-$p$ hybridization of the states 
in the proximity of the Fermi energy. Our calculated electronic structure and 
optical spectra are in good 
agreement with experiments. We obtain a band gap of 1.25 eV 
and local moment per Cu atom is 0.68 $\mu_B$ with 
significant magnetic moment of O $\sim$ 0.18 $\mu_B$. The comparison of our calculated 
electronic structure with PES show generally good agreement with the key binding energies 
correctly obtained in our computation. So is our computed optical spectra.    
\label{sec:conclusion}

{\bf Acknowledgments.}
Work at LSU is funded in part by the National Science Foundation award No. EPS-1003897. V. I. Anisimov 
acknowledgments the supports of the following. The Russian Foundation for Basic Research 
(Projects No. 13-02-00050 and No. 12-02-91371-CT$_a$), the fund of the President of the 
Russian Federation for the support of scientific schools NSH-6172.2012.2, 
the Program of the Russian Academy of Science Presidium ``Quantum microphysics of condensed 
matter'' 12-P-2-1017, and the grant of the Ministry of education and science of Russia No. 14.A18.21.0076. 
High performance computational resources were provided by Louisiana Optical Network Initiative (LONI).
C. E. Ekuma wishes to thank the Government of Ebonyi State, Nigeria.


\begin{thebibliography}{10}

\bibitem{Ghijsen1988}
J.~Ghijsen, L.~H. Tjeng, J.~van Elp, H.~Eskes, J.~Westerink, G.~A. Sawatzky,
  and M.~T. Czyzyk, Phys. Rev. B {\bf 38}, 11322--11330 (1988).

\bibitem{0295-5075-7-6-011}
F.~Parmigiani and G.~Samoggia,  Europys. Lett. {\bf 7}(6), 543 (1988).

\bibitem{PhysRevLett.85.5170}
X.~G. Zheng, C.~N. Xu, Y.~Tomokiyo, E.~Tanaka, H.~Yamada, and Y.~Soejima, Phys.
  Rev. Lett. {\bf 85}, 5170--5173 (2000).

\bibitem{RevModPhys.60.585}
J.~G. Bednorz and K.~A. M\"uller, Rev. Mod. Phys. {\bf 60}, 585--600 (1988).

\bibitem{Bednorz1986}
J.~Bednorz and K.~M\"{u}ller, Z. Physik B Conden. Matter {\bf 64}(2),
  189--193 (1986).

\bibitem{PhysRevB.5.290}
L.~F. Mattheiss, Phys. Rev. B {\bf 5}, 290--306 (1972).

\bibitem{PhysRevB.30.4734}
K.~Terakura, T.~Oguchi, A.~R. Williams, and J.~K\"ubler, Phys. Rev. B {\bf 30},
  4734--4747 (1984).

\bibitem{PhysRevB.76.054417}
W.~A. Harrison, Phys. Rev. B {\bf 76}, 054417 (2007).

\bibitem{PhysRevB.68.224433}
T.~Shimizu, T.~Matsumoto, A.~Goto, T.~V. Chandrasekhar~Rao, K.~Yoshimura, and
  K.~Kosuge, Phys. Rev. B {\bf 68}, 224433 (2003).

\bibitem{PhysRevB.39.4343}
B.~X. Yang, T.~R. Thurston, J.~M. Tranquada, and G.~Shirane, Phys. Rev. B {\bf
  39}, 4343--4349 (1989).

\bibitem{0022-3719-21-15-023}
J.~B. Forsyth, P.~J. Brown, and B.~M. Wanklyn, J. Phys. C: Solid State Physics
  {\bf 21}(15), 2917 (1988).

\bibitem{Anisimov1997}
V.~I. Anisimov, F.~Aryasetiawan, and A.~I. Lichtenstein, J. Phys.: Condens.
  Matter {\bf 9}, 767 -- 808 (1997).

\bibitem{PhysRevB.44.943}
V.~I. Anisimov, J.~Zaanen, and O.~K. Andersen, Phys. Rev. B {\bf 44}, 943--954
  (1991).

\bibitem{Poizot2000}
P.~Poizot, S.~Laruelle, S.~Grugeon, and J.~M. Taracon, Nature {\bf 407}, 496
  (2000).

\bibitem{Ray2001}
S.~C. Ray, Sol. Energy Mater. Sol. Cells {\bf 68}, 307 (2001).

\bibitem{Ben2009}
T.~Ben-Moshe, I.~Dror, and B.~Berkowitz, Appl. Catal. B {\bf 85}, 207--211
  (2009).

\bibitem{A801595C}
T.~Ishihara, M.~Higuchi, T.~Takagi, M.~Ito, H.~Nishiguchi, and Y.~Takita, J.
  Mater. Chem. {\bf 8}, 2037--2042 (1998).

\bibitem{Jiang1998}
Y.~Jiang, S.~Decker, C.~Mohs, and K.~J. Klabunde, J. Catal. {\bf 180}, 24
  (1998).

\bibitem{Morales2009}
J.~Morales, L.~S\'{a}nchez, F.~Mart\'{i}n, R.-B.~J. R, and M.~S\'{a}nchez,
  Electrochim Acta {\bf 49}, 4589--459 (2004).

\bibitem{doi:10.1021/cm000166z}
R.~V. Kumar, Y.~Diamant, and A.~Gedanken, Chemistry of Materials {\bf 12}(8),
  2301--2305 (2000).

\bibitem{PhysRevB.61.11093}
K.~Borgohain, J.~B. Singh, M.~V. Rama~Rao, T.~Shripathi, and S.~Mahamuni, Phys.
  Rev. B {\bf 61}, 11093--11096 (2000).

\bibitem{Rehman2011}
S.~Rehman, A.~Mumtaz, and S.~K. Hasanain, J. Nanopart. Res. {\bf 13},
  2497--2507 (2011).

\bibitem{Kimura2008}
T.~Kimura, Y.~Sekio, H.~Nakamura, T.~Siegrist, and A.~P.Ramirez, Nature Mater.
  {\bf 7}, 291 (2008).

\bibitem{PhysRevLett.108.187205}
G.~Jin, K.~Cao, G.-C. Guo, and L.~He, Phys. Rev. Lett. {\bf 108}, 187205
  (2012).

\bibitem{Samokhvalov1989}
A.~A. Samokhvalov, N.~N. Loshkareva, Y.~P. Sukhorukov, V.~A. Gruverman, B.~A.
  Gizhveski\u{i}, and N.~M. Chebotaev, JETP Lett. {\bf 49}, 523 (1989).

\bibitem{ICSD2011}
ICSD, {\em Inorganic Crystal Structure Database (ICSD), National Institute of
  Standards and Technology (NIST) Release 2013/1\/}, vol.~1 (NIST, 2013).

\bibitem{0953-8984-24-17-175002}
D.~Tahir and S.~Tougaard, J. Phys.: Conden. Matter {\bf 24}(17), 175002
  (2012).

\bibitem{PhysRevB.52.1433}
F.~Marabelli, G.~B. Parravicini, and F.~Salghetti-Drioli, Phys. Rev. B {\bf
  52}, 1433--1436 (1995).

\bibitem{Madelung2006}
O.~Madelung, U.~R\"{o}ssler, and M.~Schulz (eds.) {\em Numerical Data and
  Functional Relationships in Science and Technology, Landolt-B\"{o}rnstein,
  New Series, Group III\/}, vol. 17a, and 22a (Springer, 2006).

\bibitem{0953-8984-4-23-009}
M.~Ain, A.~Menelle, B.~M. Wanklyn, and E.~F. Bertaut, J. Phys.: Conden. Matter {\bf 4}(23), 5327 (1992).

\bibitem{PhysRevB.40.7684}
W.~Y. Ching, Y.-N. Xu, and K.~W. Wong, Phys. Rev. B {\bf 40}, 7684--7695
  (1989).

\bibitem{Kanagaraj}
C.~Kanagaraj and N.~Baskaran, Advanced Materials Research {\bf 488--489}, 129
  (2012).

\bibitem{PhysRevB.87.115111}
M.~Heinemann, B.~Eifert, and C.~Heiliger, Phys. Rev. B {\bf 87}, 115111 (2013).

\bibitem{PhysRevB.49.14211}
M.~T. Czy\ifmmode~\dot{z}\else \.{z}\fi{}yk and G.~A. Sawatzky, Phys. Rev. B
  {\bf 49}, 14211--14228 (1994).

\bibitem{PhysRevB.52.R5467}
A.~I. Liechtenstein, V.~I. Anisimov, and J.~Zaanen, Phys. Rev. B {\bf 52},
  R5467--R5470 (1995).

\bibitem{PhysRevB.73.235206}
D.~Wu, Q.~Zhang, and M.~Tao, Phys. Rev. B {\bf 73}, 235206 (2006).

\bibitem{0953-8984-22-4-045502}
X.~Rocquefelte, M.-H. Whangbo, A.~Villesuzanne, S.~Jobic, F.~Tran, K.~Schwarz,
  and P.~Blaha, J. Phys.: Conden. Matter {\bf 22}(4), 045502 (2010).

\bibitem{PhysRevB.79.195122}
W.~Siemons, G.~Koster, D.~H.~A. Blank, R.~H. Hammond, T.~H. Geballe, and M.~R.
  Beasley, Phys. Rev. B {\bf 79}, 195122 (2009).

\bibitem{PhysRevLett.106.026401}
G.~Giovannetti, S.~Kumar, A.~Stroppa, J.~van~den Brink, S.~Picozzi, and
  J.~Lorenzana, Phys. Rev. Lett. {\bf 106}, 026401 (2011).

\bibitem{PhysRevLett.106.257601}
P.~Tol\'edano, N.~Leo, D.~D. Khalyavin, L.~C. Chapon, T.~Hoffmann, D.~Meier,
  and M.~Fiebig, Phys. Rev. Lett. {\bf 106}, 257601 (2011).

\bibitem{PhysRevB.84.115108}
B.~Himmetoglu, R.~M. Wentzcovitch, and M.~Cococcioni, Phys. Rev. B {\bf 84},
  115108 (2011).

\bibitem{Gizhevski2006}
B.~A. Gizhevski\u{i}, A.~S.~M. Yu. P.~Sukhorukov, N.~N. Loshkareva, E.~V.
  Mostovshchikova, A.~E. Ermakov, E.~A. Kozlov, M.~A. U\u{i}min, and V.~S.
  Gaviko, J. Exp. and Theor.Phys. {\bf 102}, 298 (2006).

\bibitem{0953-8984-17-3-009}
B.~A. Gizhevskii, Y.~P. Sukhorukov, N.~N. Loshkareva, A.~S. Moskvin, E.~V.
  Zenkov, and E.~A. Kozlov, J. Phys.: Conden. Matter {\bf 17}(3), 499 (2005).

\bibitem{0953-8984-11-26-306}
S.~Warren, W.~R. Flavell, A.~G. Thomas, J.~Hollingworth, P.~L. Wincott, A.~F.
  Prime, S.~Downes, and C.~Chen, J. Phys.: Conden. Matter {\bf 11}(26), 5021
  (1999).

\bibitem{singh2006}
D.~J. Singh, {\em Planewaves, Pseudopotentials, and the LAPW Method, 2nd Ed.\/}
  (Springer-Velag, Berlin, 2006).

\bibitem{Blaha2001}
P.~Blaha, K.~Schwarz, G.~Madsen, D.~Kvasnicka, and J.~Luitz, {\em WIEN2K, An
  Augmented Plane Wave+Local Orbitals Program for Calculating Crystal
  Structure\/} (K. Schwarz Technical University, Wien, Austria, 2001).

\bibitem{PhysRevB.71.035105}
M.~Cococcioni and S.~de~Gironcoli, Phys. Rev. B {\bf 71}, 035105 (2005).

\bibitem{Anisimov1991}
V.~I. Anisimov and O.~Gunnarsson, Phys. Rev. B {\bf 43}, 7570 (1991).

\bibitem{Madsen2005}
G.~K.~H. Madsen and P.~Novark, Europys. Lett. {\bf 69}, 777 (2005).

\bibitem{Perdew1996}
J.~P. Perdew, K.~Burke, and M.~Ernzerhof, Phys. Rev. Lett. {\bf 77}, 3865
  (1996).

\bibitem{PhysRevB.87.085112}
S.~Lany, Phys. Rev. B {\bf 87}, 085112 (2013).

\bibitem{Shen1990}
Z.~X. Shen, R.~S. List, D.~S. Dessau, F.~Parmigiani, A.~J. Arko, R.~Bartlett,
  B.~O. Wells, I.~Lindau, and W.~E. Spicer, Phys. Rev. B {\bf 42}, 8081 (1990).

\bibitem{Thuler1982}
M.~R. Thuler, R.~L. Benbow, and Z.~Hurych, Phys. Rev. B {\bf 26}, 669 (1982).

\bibitem{PhysRevLett.78.1126}
L.~H. Tjeng, B.~Sinkovic, N.~B. Brookes, J.~B. Goedkoop, R.~Hesper,
  E.~Pellegrin, F.~M.~F. de~Groot, S.~Altieri, S.~L. Hulbert, E.~Shekel, and
  G.~A. Sawatzky, Phys. Rev. Lett. {\bf 78}, 1126--1129 (1997).

\bibitem{Ghijsen1990}
J.~Ghijsen, L.~H. Tjeng, H.~Eskes, G.~A. Sawatzky, and R.~L. Johnson, Phys.
  Rev. B {\bf 42}, 2268--2274 (1990).

\bibitem{0965-0393-14-2-006}
M.~A. M\"{a}ki-Jaskari, Modelling and Simulation in Materials Science and
  Engineering {\bf 14}(2), 207 (2006).

\bibitem{PhysRevB.56.12818}
M.~Takahashi and J.-i. Igarashi, Phys. Rev. B {\bf 56}, 12818--12824 (1997).

\bibitem{PhysRevB.41.288}
H.~Eskes, L.~H. Tjeng, and G.~A. Sawatzky, Phys. Rev. B {\bf 41}, 288--299
  (1990).

\bibitem{PhysRevB.38.6650}
A.~K. McMahan, R.~M. Martin, and S.~Satpathy, Phys. Rev. B {\bf 38}, 6650--6666
  (1988).

\bibitem{Harold1966}
H.~Wieder and A.~W. Czanderna, J. Appl. Phys. {\bf 37}, 184 (1966).

\bibitem{Jundale}
D.~Jundale, P.~Joshi, S.~Sen, and V.~Patil, J. Mater. Sci.: Materials in
  Electronics {\bf 23}(8), 1492--1499 (2012).

\bibitem{Chen2009927}
A.~Chen, H.~Long, X.~Li, Y.~Li, G.~Yang, and P.~Lu, Vacuum {\bf 83}(6), 927 --
  930 (2009).

\bibitem{Balamurugan200190}
B.~Balamurugan and B.~Mehta, Thin Solid Films {\bf 396}(1–2), 90 -- 96
  (2001).

\bibitem{Aspnes1985}
D.~E. Aspnes, {\em in Handbook of Optical Constants of Solids\/} (Academic, New
  York, 1985).

\bibitem{Ito1998}
T.~Ito, , H.~Yamaguchi, T.~Masumi, and S.~Adachi, J. Phys. Soc. Jpn. {\bf 67},
  3304--3309 (1998).

\bibitem{PSSB:PSSB201248128}
B.~K. Meyer, A.~Polity, D.~Reppin, M.~Becker, P.~Hering, P.~J. Klar, T.~Sander,
  C.~Reindl, J.~Benz, M.~Eickhoff, C.~Heiliger, M.~Heinemann, J.~Bl\"{a}sing,
  A.~Krost, S.~Shokovets, C.~Müller, and C.~Ronning, physica status solidi (b)
  {\bf 249}(8), 1487--1509 (2012).

\bibitem{Nomerovannaya}
L.~V. Nomerovannaya, A.~A. Makhnev, M.~M. Kirillova, S.~V. Moshkin, M.~G.
  Lyubimov, and N.~V. Egorova, Superconductivity {\bf 3}, 159 (1990).

\bibitem{PhysRevB.63.094303}
A.~B. Kuz'menko, D.~van~der Marel, P.~J.~M. van Bentum, E.~A. Tishchenko,
  C.~Presura, and A.~A. Bush, Phys. Rev. B {\bf 63}, 094303 (2001).

\bibitem{PhysRevB.51.3140}
C.~C. Homes, M.~Ziaei, B.~P. Clayman, J.~C. Irwin, and J.~P. Franck, Phys. Rev.
  B {\bf 51}, 3140--3150 (1995).

\end{thebibliography}

\end{document}